\documentclass[preprint,onecolumn,amsmath,amssymb,floatfix]{revtex4}
\usepackage{graphicx}

\begin{document}

\title{Electromigrated nanoscale gaps for surface-enhanced Raman spectroscopy}

\author {Daniel R. Ward$^{1}$, Nathaniel K. Grady$^{2}$, Carly S. Levin$^{3}$,  Naomi J. Halas$^{3,4}$, Yanpeng Wu$^{2}$, Peter Nordlander$^{1,4}$, Douglas Natelson$^{1,4}$}
\affiliation{$^{1}$Department of Physics and Astronomy, 
$^{2}$Applied Physics Graduate Program,
$^{3}$Department of Chemistry,
$^{4}$Department of Electrical and Computer Engineering, and the Rice Quantum Institute, Rice University, 6100 Main St., Houston, TX  77005, USA}

% Include the date command, but leave its argument blank.

\date{\today}

\begin{abstract}
Single-molecule detection with chemical specificity is a powerful and
much desired tool for biology, chemistry, physics, and sensing
technologies.  Surface-enhanced spectroscopies enable single molecule
studies, yet reliable substrates of adequate sensitivity are in short
supply.  We present a simple, scaleable substrate for surface-enhanced
Raman spectroscopy (SERS) incorporating nanometer-scale
electromigrated gaps between extended electrodes.  Molecules in the
nanogap active regions exhibit hallmarks of very high Raman
sensitivity, including blinking and spectral diffusion.  Electrodynamic
simulations show plasmonic focusing, giving electromagnetic
enhancements approaching those needed for single-molecule SERS.
\end{abstract}

\maketitle

Multifunctional sensors with single-molecule sensitivity are greatly
desired for a variety of sensing applications, from biochemical
analysis to explosives detection.  Chemical and electromagnetic
interactions between molecules and metal substrates are used in
surface-enhanced spectroscopies\cite{Moskovits85RMP} to approach
single molecule sensitivity.  Electromagnetic enhancement in
nanostructured conductors results when incident light excites local
electronic modes, producing large electric fields in a nanoscale
region, known as a ``hot spot", that greatly exceed the strength of
the incident field.  Hot spots can lead to particularly large
enhancements of Raman scattering, since the Raman scattering rate is
proportional to $|{\bf E}(\omega)|^{2}|{\bf E}(\omega')|^{2}$ at the
location of the molecule, where ${\bf E}(\omega)$ is the electric
field component at the frequency of the incident radiation, and ${\bf
E}(\omega')$ is the component at the scattered frequency.

It has been an ongoing challenge to design and fabricate a substrate
for systematic SERS at the single molecule level.  Single-molecule SERS sensitivity was first clearly demonstrated using
random aggregates of colloidal
nanoparticles\cite{KneippetAl97PRL,NieetAl97Science,XuetAl99PRL,MichaelsetAl00JPCB}.
Numerous other metal substrate configurations have been used for SERS,
including chemically engineered
nanoparticles\cite{JacksonetAl04PNAS,WangetAl05JACS,OldenburgetAl99JCP}, nanostructures
defined by bottom-up patterning\cite{HaynesetAl01JPCB,QinetAl06PNAS},
and those made by traditional lithographic
approaches\cite{FrommetAl06JCP}.  In the most sensitive substrate
geometries, incident light excites adjacent subwavelength nanoparticles or nanostructures, resulting in large field enhancements within the interparticle gap\cite{HallocketAl05PNAS,NordlanderetAl04NL}.  Fractal aggregates of
nanoparticles\cite{WangetAl03PNAS} can further increase field
enhancements by focusing plasmon energy from larger length scales down
to particular nanometer-scale hotspots\cite{LietAl03PRL}.  However,
precise and reproducible formation of such assemblies in
predetermined locations has been extremely challenging.  An
alternative approach is tip-enhanced Raman spectroscopy (TERS), in
which the incident light excites an interelectrode plasmon resonance
localized between a sharp, metal scanned probe tip and an underlying
metal substrate.  Recent progress has been made in single-molecule TERS detection\cite{DomkeetAl06JACS,NeascuetAl06PRB,ZhangetAl07JPCC}.  A similar
approach was recently attempted using a mechanical break
junction\cite{TianetAl06JACS}.  While useful for surface imaging, TERS
requires feedback to control the tip-surface gap, and is not scalable
or readily integrated with other sensing modalities.

We demonstrate a scaleable and highly reliable method for producing
planar extended electrodes with nanoscale spacings that exhibit very
large SERS signals, with each electrode pair having one well-defined
hot spot.  Confocal scanning Raman microscopy demonstrates the
localization of the enhanced Raman emission.  The SERS response is
consistent with a very small number of molecules in the hotspot,
showing blinking and spectral diffusion of Raman lines.  Sensitivity is
sufficiently high that SERS from physisorbed atmospheric contaminants
may be detected after minutes of exposure to ambient conditions.  The
Raman enhancement for $para$-mercaptoaniline ($p$MA) is estimated from
experimental data to exceed $10^{8}$.  Finite-difference time-domain
(FDTD) modeling of realistic structures reveals a rich collection of
interelectrode plasmon modes that can readily lead to SERS
enhancements as large as $5 \times 10^{10}$ over a broad range of
illumination wavelengths.  These structures hold the promise of
integration of single-molecule SERS with electronic transport
measurements, as well as other near-field optical devices.

%Figure 1:
%Fabrication, showing electron micrograph of unbroken multibowtie
%structure, close-up of broken region after electromigration,
%maps of Si, Rayleigh (Au), and Raman signature.

\begin{figure}[h!]
\begin{center}
\includegraphics[clip, width=8cm]{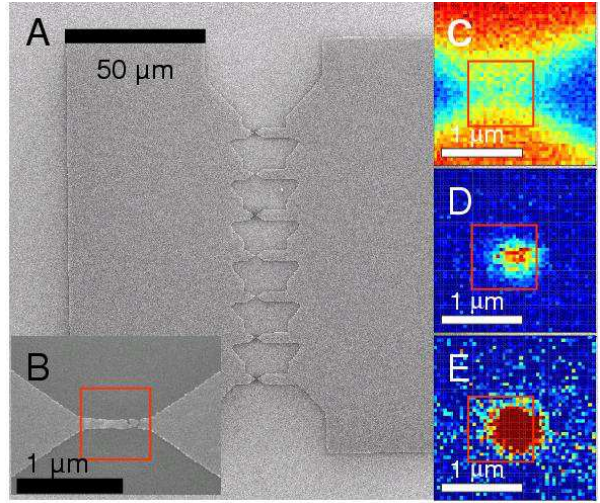}
\end{center}
%\vspace{-5mm}
\caption{\small (A) Full multibowtie structure, with seven
nanoconstrictions. (B) Close-up of an individual constriction after
electromigration.  Note that the resulting nanoscale gap ($< \sim$~5~nm
at closest separation, as inferred from closer images) is toward the
right edge of the indicated red square. (C) Map of Si Raman peak
(integrated from 500-550~cm$^{-1}$) in device from (B), with red corresponding
to high total counts.  The attenuation of the Si Raman
line by the Au electrodes is clear.  (D) Map of $p$MA SERS signal
for this device based on one carbon ring mode (integrated from 1050-1110~cm$^{-1}$).  (E) Map of integrated low energy background (50-300~cm$^{-1}$) 
for this device.}
\label{fig1}
%\vspace{-5mm}
\end{figure}

Our structures are fabricated on a Si wafer topped by 200~nm of
thermal oxide.  Electron beam lithography is used to pattern
``multibowtie'' structures as shown in Fig.~\ref{fig1}A.  The
multibowties consist of two larger pads connected by multiple
constrictions, as shown.  The constriction widths are 80-100~nm,
readily within the reach of modern photolithography.  After
evaporation of 1~nm Ti and 15~nm Au followed by liftoff in acetone,
the electrode sets are cleaned of organic residue by exposure to
O$_{2}$ plasma for 1~minute.  The multibowties are placed in a vacuum
probe station, and electromigration\cite{ParketAl99APL} is used to
form nanometer-scale gaps in the constrictions in parallel, as shown
in Fig.~1B.  Electromigration is a nonthermal process whereby momentum
transfer from current-carrying electrons is transferred to the
lattice, rearranging the atomic positions.  Electromigration has been
studied thoroughly\cite{StrachanetAl05APL,TrouwborstetAl06JAP,StrachanetAl06NL,TaychatanapatetAl07NL} as a means of producing electrodes for
studies of single-molecule conduction.  We have performed manual and
automated electromigration at room temperature, with identical
results.  The number of parallel constrictions in a single multibowtie
is limited by the output current capacity of our electromigration
voltage source.  A post-migration resistance of $\sim$~10~k$\Omega$
for the structure in Fig.~1A appears optimal.

Post-migration high resolution scanning electron microscopy (SEM)
shows interelectrode gaps ranging from too small to resolve to several
nanometers.  There are no detectable nanoparticles in the gap region
or along the electrode edges.  Based on electromigration of 283
multibowties (1981 individual constrictions), 77\% of multibowties
have final resistances less than 100~k$\Omega$, and 43\% have final
resistances less than 25~k$\Omega$.  We believe that this yield,
already high, can be improved significantly with better process
control, particularly of the lithography and liftoff.

The optical properties of the resulting nanogaps are characterized
using a WITec CRM~200 scanning confocal Raman microscope in reflection
mode, with normal illumination from a 785~nm diode laser.  Using a
100$\times$ objective, the resulting diffraction-limited spot is
measured to be gaussian with a full-width at half-maximum of 575~nm.
Fig.~\ref{fig1}C shows a spatial map of the integrated emission from the
520~cm$^{-1}$ Raman line of the Si substrate.  The Au electrodes are
clearly resolved.  Polarization of the incident radiation is horizontal
in this figure.  Rayleigh scattered light from these structures
shows significant changes upon polarization rotation, while SERS
response is approximately {\it independent} of polarization. 

Freshly cleaned nanogaps show no Stokes-shifted Raman emission out to
3000~cm$^{-1}$.  However, in 65\% of clean nanogaps examined, a broad
continuum background (see Supporting Information) is seen, decaying
roughly linearly in wavenumber out to 1000~cm$^{-1}$ before falling
below detectability.  This background is spatially localized to a
diffraction-limited region around the interelectrode gap and is
entirely absent in unmigrated junctions.  The origin of this
continuum, similar to that seen in other strongly enhancing SERS
substrates\cite{MichaelsetAl00JPCB}, is likely inelastic
electronic effects in the gold electrodes\cite{BeversluisetAl03PRB}.
In samples coated with molecules, this background correlates strongly
with visibility of SERS.  No junctions without this background displayed 
SERS signals.  Like the SERS signal, this background is approximately
polarization independent.  Temporal fluctuations of this background in
clean junctions are minimal, strongly implying that fluctuations of
the electrode geometry are not responsible for SERS blinking.

The SERS enhancement of the junctions has been tested using various
molecules.  The bulk of testing utilized $p$MA, 
which self-assembles onto the Au electrodes via standard thiol
chemistry.  Particular modes of interest are carbon ring modes at
1077~cm$^{-1}$ and 1590~cm$^{-1}$.  Fig.~\ref{fig1}D shows a map of the Raman
emission from the 1077~cm$^{-1}$ line on the same junction as
Fig.~\ref{fig1}B,C, after self-assembly of $p$MA.  This emission is strongly
localized to the position of the nanogap.  No Raman signal is
detectable either on the metal films or at the edges of the metal
electrodes.  Fig.~\ref{fig1}E shows the spatial localization of the continuum
background mentioned above.

\begin{figure}[h!]
\begin{center}
\includegraphics[clip, width=8cm]{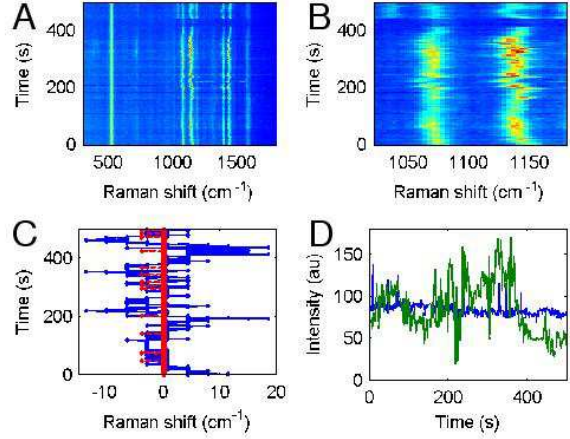}
\end{center}
%\vspace{-5mm}
\caption{\small (A) Waterfall plot (1~s integration steps) of SERS
spectrum at a single nanogap that had been soaked in 1~mM $p$MA in ethanol.
Identified $p$MA peaks include the ring modes at 1077~cm$^{-1}$ and
1590~cm$^{-1}$, as well as an 1145~cm$^{-1}$ $\delta$CH mode with
b$_{2}$ symmetry, an 1190~cm$^{-1}$ mode identified as $\delta$CH of
a$_{1}$ symmetry, a mode near 1380~cm$^{-1}$ identified as
$\delta$CH+$\nu$CC of b$_{2}$ symmetry, and a mode near 1425~cm$^{-1}$
identified as $\nu$CC+$\delta$CH of b$_{2}$ symmetry.  Mode
assignments are based on \protect{Ref.~\cite{OsawaetAl94JPC}}.  (B) Close-up of (A)
to show correlated wandering and blinking of 1077~cm$^{-1}$ and 1145~cm$^{-1}$
modes.  (C) Comparison of 1145~cm$^{-1}$ mode position (blue) with the Si Raman peak (red), which shows no such wandering.  The jitter in the Si peak position is 1 pixel in the detector. (D) Comparison of 1145~cm$^{-1}$ peak height (green, found by a gaussian fit to the peak) fluctuations with those of the Si peak (blue).  }
\label{fig2}
%\vspace{-5mm}
\end{figure}

Fig.~\ref{fig2} shows a more detailed examination of the SERS response of the
gap region of a typical junction after self-assembly of $p$MA.
Fig.~\ref{fig2}A,B are time series of the SERS response, with known $p$MA modes
indicated.  The modes visible are similar to those seen in other SERS
measurements of $p$MA on lithographically fabricated Au
structures\cite{FrommetAl06JCP}.  Each spectrum was acquired with 1~s
integration time, with the objective positioned over the center of
the nanogap hotspot.  Temporal fluctuations of both the Raman
intensity (``blinking'') and Raman shift (spectral diffusion), generally
regarded as hallmarks of few- or single-molecule SERS
sensitivity\cite{WangetAl05JPCB}, are readily apparent.  Fig.~\ref{fig2}C shows
a comparison of the wandering of the 1077~cm$^{-1}$ $p$MA line with that
of the 520~cm$^{-1}$ Raman line of the underlying Si substrate over
the same time interval.  This demonstrates that the spectral diffusion is due
to changes in the molecular environment, rather than a variation in
spectrometer response.  Fig.~\ref{fig2}D shows the variation in the peak
amplitudes over that same time interval.

This blinking and spectral diffusion are seen {\it routinely} in these
junctions.  We have observed such Raman response from several
molecules, including self-assembled films of $p$MA,
$para$-mercaptobenzoic acid ($p$MBA), a Co-containing transition metal
complex\cite{CiszeketAl06JACS}, and spin-coated poly(3-hexylthiophene)
(P3HT).  These molecules all have distinct Raman responses and show
blinking and wandering in the junction hotspots.

Another indicator of very large enhancement factors in these
structures is sensitivity to exogenous, physisorbed contamination.
Carbon contamination has been
discussed\cite{KudelskietAl00CPL,Otto02JRS,RichardsetAl03JRS} in the context of both
SERS and TERS.  This substrate is sensitive enough to examine such
contaminants (see Supporting Information).  While clean junctions with
no deliberately attached molecules initially show only the continuum
background, gap-localized SERS signatures in the $sp^{2}$ carbon
region between 1000~cm$^{-1}$ and 1600~cm$^{-1}$ are readily detected
after exposure to ambient lab conditions for tens of minutes.
Nanojunctions that have been coated with a self-assembled monolayer
(SAM) (for example, $p$MA) do {\it not} show this carbon signature,
even after hours of ambient exposure.  Presumably this has to do with
the extremely localized field enhancement in these structures, with
the SAM sterically preventing physisorbed contaminants from entering
the region of enhanced near field.

Recently arrived contaminant SERS signatures abruptly disappear within
tens of seconds at high incident powers (1.8~mW), presumably due to
desorption.  SERS from covalently bound molecules is considerably more
robust, degrading slowly at high powers, and persisting indefinitely
for incident powers below 700~$\mu$W.  SEM examination of the nanogaps
shows no optically induced damage after exposure to intensities that
would significantly degrade nanoparticles\cite{SchucketAl05PRL}.  The
large extended pads likely aid in the thermal sinking of the nanogap
region to the substrate.

\begin{figure}[h!]
\begin{center}
\includegraphics[clip, width=8cm]{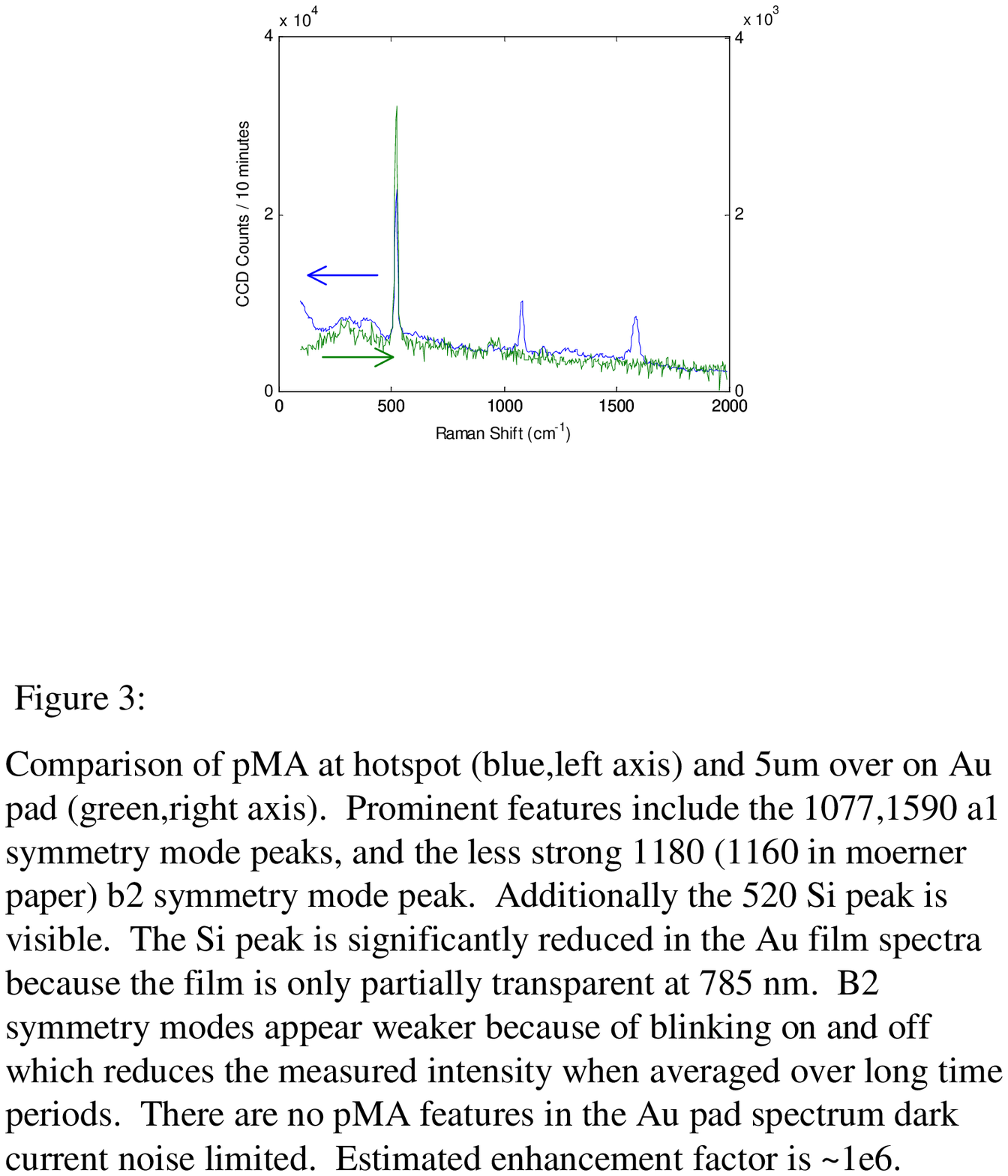}
\end{center}
%\vspace{-5mm}
\caption{\small Blue curve (left scale): $p$MA SERS spectrum at
hotspot center of one nanojunction densely covered by $p$MA,
integrated for 10 minutes at incident power of 700~$\mu$W. Green curve
(right scale): integrated signal under same conditions on middle of Au
pad on the same nanojunction.  The feature near 960~cm$^{-1}$ is from
the Si substrate.  No Raman features are detectable on either the Au
pads or their edges under these conditions.  }
\label{fig3}
%\vspace{-5mm}
\end{figure}

Estimating enhancement factors rigorously is notoriously difficult,
particularly when the hotspot size is not known.  Although SERS
enhancement volume measurements are possible using molecular
rulers\cite{LaletAl06NL}, this is not feasible with such small
nanogaps.  Junctions made directly on Raman-active substrates (Si with
no oxide; GaAs) show no clearly detectable enhancement of substrate
modes in the gap region, suggesting that the electromagnetic
enhancement is strongly confined to the thickness of the metal film
electrodes.  Figure~\ref{fig3} shows a comparison between a typical $p$MA SERS
spectrum acquired on a junction with a 600~s integration time at
700~$\mu$W incident power, and a spectrum acquired over one of that
device's Au pads, for the same settings.  The pad spectrum shows no
detectable $p$MA features and is dark current limited.

\begin{figure}[h!]
\begin{center}
\includegraphics[clip, width=12.5cm]{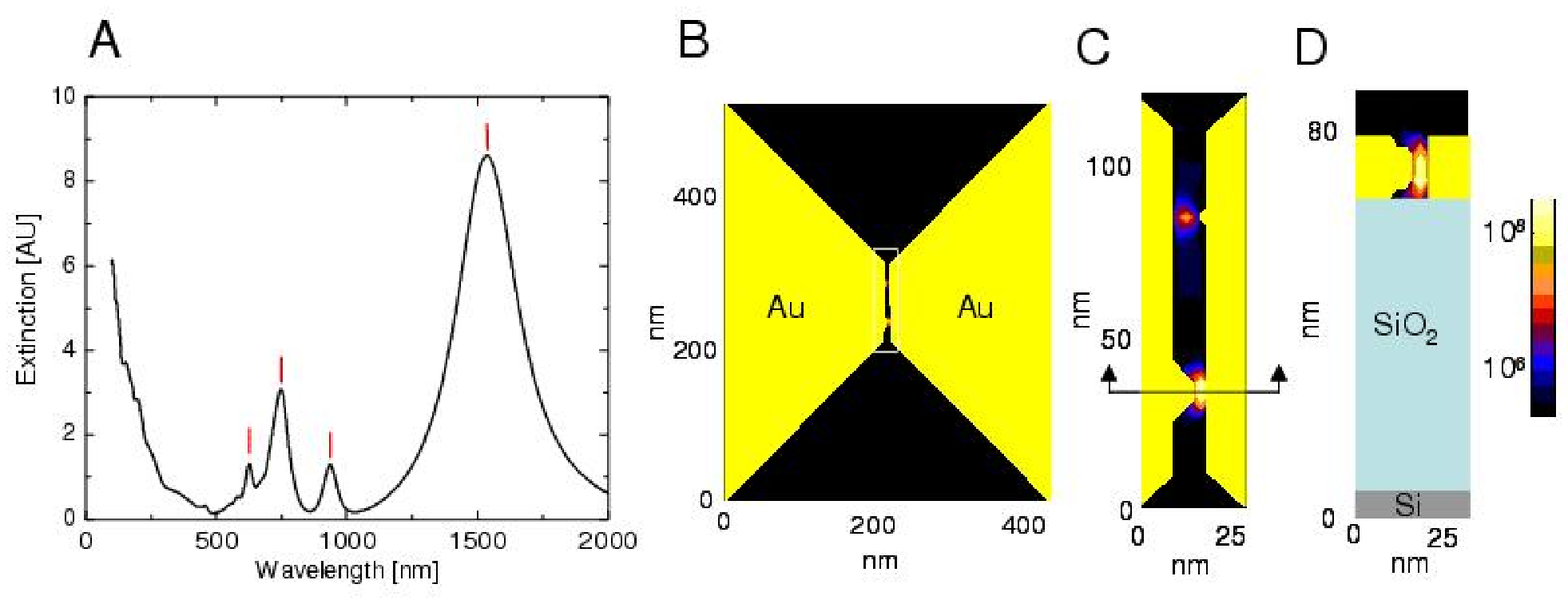}
\end{center}
%\vspace{-5mm}
\caption{\small (A) FDTD-calculated extinction spectrum from the
model electrode configuration shown in (B).  
(B) Mock-up electrode tips capped with nanoscale hemispherical asperities,
with $|{\mathbf E}|^{4}$ plotted for the 937~nm resonance of (A).
Constriction transverse width at narrowest point is 100~nm.  Gap size
without asperities is 8~nm.  Asperity on left (right) electrode has
radius of 6~nm (4~nm).  Au film thickness is 15~nm, SiO$_{2}$
underlayer thickness is 50~nm.  Radiation is normally incident,
with polarization oriented horizontally.  Grid size for FDTD
calculation is 2~nm.  (C) Close-up of central region of (B),
showing extremely localized enhancement at asperities.  (D) Cross-section
indicated in (C), showing that enhancement in this configuration
does not penetrate significantly into the substrate.  Predicted
maximum electromagnetic Raman enhancement in this mode exceeds $10^{8}$.
}
\label{fig4}
%\vspace{-5mm}
\end{figure}

We use FDTD calculations to understand the strong SERS response in
these structures and roughly estimate enhancement factors.  It is
important to note that the finite grid size (2~nm) required for
practical computation times restricts the quantitative accuracy of
these calculations.  However, the main results regarding spatial mode
structure (allowing assessment of the localization of the hotspot) and
energy dependence are robust to these concerns, and the calculated
electric field enhancement is an {\it
underestimate}\cite{OubreNordlander05JPCB}.  Fig.~\ref{fig4} shows a
calculated extinction spectrum and map of $|{\mathbf E}|^{4}$ in the
vicinity of the junction.  Calculational details and additional plots
are presented in the Supporting Information.  These calculations
predict that there should be large SERS enhancements across a broad
bandwidth of exciting wavelengths because of the complicated mode
structure possible in the interelectrode gap.  Nanometer-scale
asperities from the electromigration process break the interelectrode
symmetry of the structure.  The result is that optical excitations at
a variety of polarizations can excite many interelectrode modes
besides the simple dipolar plasmon commonly considered.  For extended
electrodes, a continuous band of plasmon resonances coupling to
wavelengths from 500-1000~nm is expected\cite{NordlanderLe06APB}.
This broken symmetry also leads to much less dependence of the
calculated enhancement on polarization direction, as seen
experimentally.  The calculations confirm that the electromagnetic
enhancement is confined in the normal direction to the film thickness.
Laterally, the field enhancement is confined to a region comparable to
the radius of curvature of the asperity.  For gaps and asperities in
the range of 2~nm, purely electromagnetic enhancements can exceed
$10^{11}$, approaching that sufficient for single-molecule
sensitivity.

Using the data from the device in Fig.~\ref{fig3}, we estimate the total
enhancement in that device.  To be conservative, we assume a hotspot
effective radius of 2.5~nm with dense packing of $p$MA, giving $N
\approx 100$ molecules.  Blinking and wandering suggest that the true
$N$ value is much closer to one.  The integrated Raman signal over a
gaussian fit to the 1077~cm$^{-1}$ Raman line is 2.0
counts/sec/molecule when the incident power is 700~$\mu$W.  For our
apparatus the count rate from imaging a bulk crystal at the same
equivalent power (see Supporting Information) is $4.2 \times 10^{-9}$
counts/sec/molecule, so that we estimate a total enhancement of at
least $5\times 10^{8}$. 

%Conclusion and outlook.

We have demonstrated a SERS substrate capable of extremely high
sensitivity for trace chemical detection.  Unlike previous substrates,
these nanojunctions may be mass fabricated in controlled positions
with high yield using a combination of standard lithography and
electromigration.  The resulting hotspot geometry is predicted to
allow large SERS enhancements over a broad band of illuminating
wavelengths.  Other nonlinear optical effects should be observable in
these structures as well.  The extended electrode geometry and
underlying gate electrode are ideal for integration with other sensing
modalities such as electronic transport.  Tuning molecule/electrode
charge transfer via the gate electrode may also enable the direct
examination of the fundamental nature of chemical enhancement in
SERS.

DW acknowledges support from the NSF-funded Integrative Graduate
Research and Educational Training (IGERT) program in Nanophotonics.
NH, PN, and DN acknowledge support from Robert A. Welch Foundation
grants C-1220, C-1222, and C-1636, respectively.  DN also acknowledges
the National Science Foundation, the David and Lucille Packard
Foundation, the Sloan Foundation, and the Research Corporation.
C.S.L. was supported by a training fellowship from the Keck Center
Nanobiology Training Program of the Gulf Coast Consortia, NIH 1 T90
DK070121-01.  YP and NKG are supported by US Army Research Office
grant W911NF-04-1-0203.

{\bf Supporting Information Available:}  Detailed examination of 
continuum background and adsorption of exogenous contaminants; 
extended discussion of FDTD calculations; more detailed discussion
of SERS enhancement calculation.  This material is available free of 
charge via the Internet at http://pubs.acs.org.

\clearpage

\clearpage

\clearpage

\clearpage

\clearpage

{\center \bf Supporting Information:  Electromigrated nanoscale gaps for surface-enhanced Raman spectroscopy}

%\author {Daniel R. Ward$^{1}$, Nathaniel K. Grady$^{2}$, Carly S. Levin$^{3}$,  Naomi J. Halas$^{3,4}$, Yanpeng Wu$^{2}$, Peter Nordlander$^{1,4}$, Douglas Natelson$^{1,4}$}
%\affiliation{$^{1}$Department of Physics and Astronomy, 
%$^{2}$Applied Physics Graduate Program,
%$^{3}$Department of Chemistry,
%$^{4}$Department of Electrical and Computer Engineering, and the Rice Quantum Institute, Rice University, 6100 Main St., Houston, TX  77005, USA}

% Include the date command, but leave its argument blank.

%\date{\today}
%\maketitle

\renewcommand{\thefigure}{S\arabic{figure}}

\section{Continuum background}

The strong continuum background observed at the nanogaps is shown in Fig. S1A.  The continuum slopes down linearly in intensity from 0~cm$^{-1}$ to almost 1500~cm$^{-1}$.  This continuum, seen only in the presence of the nanogap, is compared with the Au film and Si substrate spectra taken using the same microscope configuration.  The 300~cm$^{-1}$ and  520~cm$^{-1}$ peaks are from the Si substrate.  The spectrum shown in Fig. S1A also shows a small peak at 1345~cm$^{-1}$ which is indicative of absorbates from the air settling at the nanogap.  The continuum is localized to the nanogap as seen in Fig. S1B, where a spatial plot has been made by integrating the SERS spectrum from 600~cm$^{-1}$ to 800~cm$^{-1}$ at each point.  This wavenumber range was chosen to avoid any of the Si substrate Raman active modes.  Additionally a comparison of Fig. S1B with the spatial plot of the Si 520~cm$^{-1}$ peak, Fig. S1C, shows that the continuum background of the nanogap is indeed located at the center of the bowtie, as expected.  Although blinking of SERS at the nanogap has been observed for $p$MA and $p$MBA, the continuum itself does not blink in the absence of molecules.  This is clear from the data in Fig S1D showing the time evolution of the SERS spectrum at the clean nanogap.  No fluctuations are observed, and the continuum background remains constant.

\begin{figure}[h!]
\begin{center}
\includegraphics[clip, width=10cm]{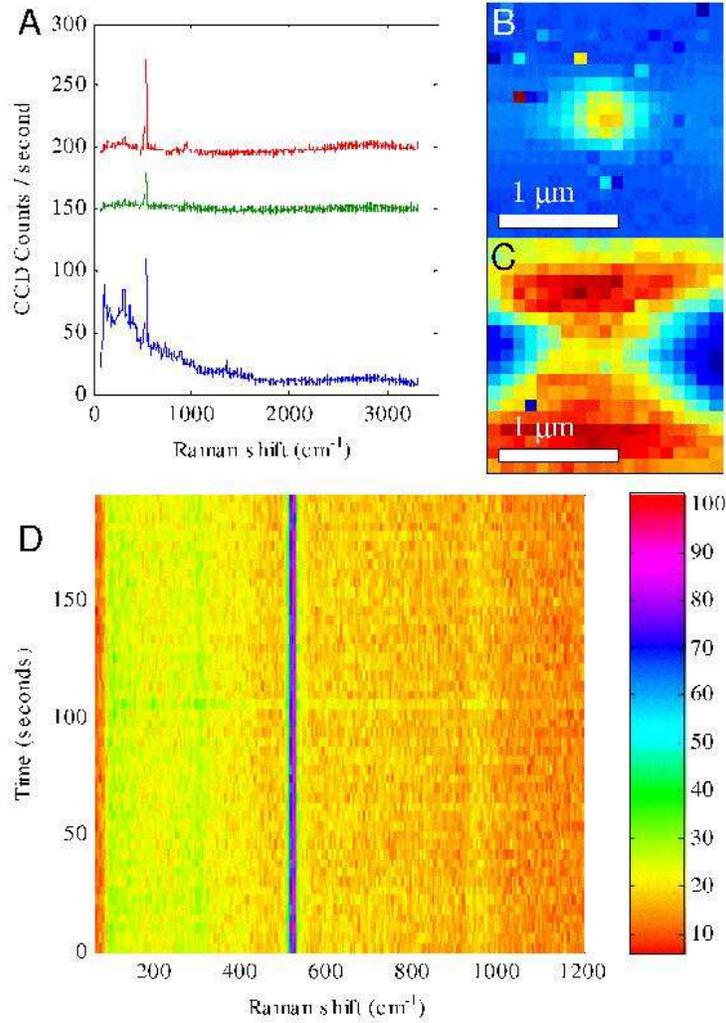}
\end{center}
\vspace{-5mm}
\caption{\small 
(A) Raman spectra at hotspot (blue) of a clean bowtie, Au pad(green), and over Si substrate(red).  The continuum is very strong at the hotspot and shows linear slope from 0 to $\sim$ 1500~cm$^{-1}$.  Also visible are the 300~cm$^{-1}$ peak and 520~cm$^{-1}$ peaks of the Si substrate and a weak peak at 1345 indicating the onset of atmospheric contamination after approximately 15 minutes of air exposure.  Curves have been offset by 150 counts/s (green) and 200 counts/s (red) for clarity.  (B) Spatial plot of integrated signal over 600-800~cm$^{-1}$ showing the localization of continuum to the center of the device when compared to the Si plot (C).  Yellow is strong signal; blue is no signal.
(C) Spatial plot of integrated Si signal over 500-540~cm$^{-1}$. Red indicates strong Si signal the blue areas show where the Au pads are.
(D) Time spectra of clean bowtie.  The intensity is reported in CCD counts/second.  No blinking of the continuum is observed.  The lowest wavenumbers are reported to have zero counts/second due to the low pass filter used to block out the laser.
}
\label{suppfig1}
%\vspace{-5mm}
\end{figure}

\clearpage

\section{Dependence on incident polarization}

The SERS  signal from the nanogap does not have significant polarization dependence, as shown in Fig S2A.  The two spectra are from the same nanogap with the polarization at 0 and 90 degrees to the gap.  Although slightly different due to positioning and actual time variation of the spectrum, the two spectra do not show any strong differences in the intensities of the $p$MA signal or the continuum.  The nanogap {\it does} exhibit a strong polarization dependence for Rayleigh scattering, as shown in Fig S2B-S2E.  Figures S2B and S2D show a spatial map of the Rayleigh scattering for the polarization across the gap (B) and parallel with the gap (D).

\begin{figure}[h!]
\begin{center}
\includegraphics[clip, width=12cm]{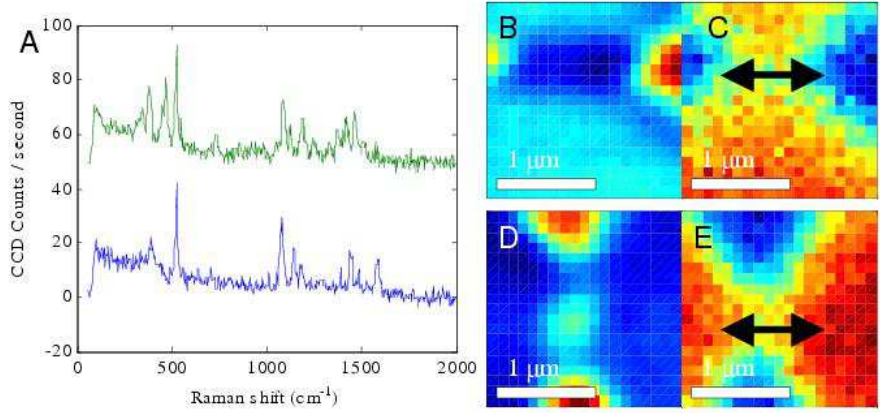}
\end{center}
%\vspace{-5mm}
\caption{\small 
(A) Raman spectra at hotspot of bowtie with $p$MA assembled on surface.  The blue spectrum is with polarization at 0 deg. (direction shown in (C)).  The green spectrum has been shifted 50 counts/s for clarity and is at the same hotspot but with the polarization rotated 90 degrees relative to the substrate (direction shown in (D)).
(B) Spatial plot of integrated signal over -40 to 40~cm$^{-1}$ showing the Rayleigh scattering from the center of the device.  Red is high intensity blue is low intensity.  The large pads are at the left/right.  Polarization direction indicated in (C).
(C) Spatial plot of integrated Si signal over 500-540~cm$^{-1}$. Red indicates Si, blue is Au pads.  Polarization direction is indicated by the arrow.
(D) Spatial plot of integrated signal over -40 to 40~cm$^{-1}$ showing the Rayleigh scattering from the center of the device for the sample rotated 90 deg. relative to (B),(C).  Red is high intensity blue is low intensity.  Polarization direction is indicated in (E).  There is a local maximum in the Rayleigh scattering at the center of the gap.
(E) Spatial plot of integrated Si signal over 500-540~cm$^{-1}$. Red indicates Si, blue is Au pads.  Polarization direction is indicated by the arrow.
}

\label{suppfig2}
%\vspace{-5mm}
\end{figure}

\clearpage

\section{Dependence on solution concentration}

We have examined SERS spectra for varying supernatant solution
concentration during the assembly procedure.  Ideally, successive
dilutions of the solution should vary surface coverage of the
assembled molecules.  While molecular coverage on the edges of
polycrystalline Au films is not readily assessed, we observe
reproducible qualitative trends as concentration is reduced.  For
$p$MBA molecules assembled from solutions in nanopure water, we have
varied concentrations from 1~mM down to 100~pM.  The fraction of
junctions showing SERS distinct from carbon contamination remains
roughly constant down to concentrations as low as 1~$\mu$M.  For our
volumes and electrode areas, this is still expected to correspond to a
dense coverage of 1 molecule per 0.19~nm$^{2}$.  At concentrations
below 1~$\mu$M, SERS spectra change significantly, while remaining
{\it distinct} from those of carbon contamination: blinking occurs
more frequently; modes of b$_{2}$ symmetry rather than a$_{1}$
symmetry appear more frequently; and the molecular peaks can be more
than 100$\times$ larger than the high coverage case for the same
integration times.  These observations are qualitatively consistent
with the molecules exploring different surface orientations at low
coverages, and charge transfer/chemical enhancement varying with
orientation.  However, the actual coverage at the edges remains
unknown.

The concentration of the solution used for assembling molecules on the
nanogap surface strongly influences the form of the observed Raman
spectrum as well as the rate and intensity of the mode blinking.
Raman spectra of $p$MBA were taken by soaking samples in 2~mL of
different concentrations of $p$MBA.  Although for all of these
concentrations there are enough molecules in solution to form a
monolayer over the bowtie surface, significant differences in the
spectra were observed.  Fig. S3A shows a representative Raman spectrum
for $p$MBA at the nanogap for 1mM concentrations.  The two carbon ring
modes at 1077~cm$^{-1}$ and 1590~cm$^{-1}$ are clearly present along
with a third peak at 1463~cm$^{-1}$.  The time spectra for this
nanogap in Fig S3B.  The 1077~cm$^{-1}$ and 1590~cm$^{-1}$ peaks are
relatively stable and always present while other modes, such as the
1463~cm$^{-1}$ mode, blink on and off for a few seconds at a time.  As
the concentration is decreased to 1~$\mu$M, the $p$MBA signal tends to
be stronger with more intense blinking.  Additionally the
1077~cm$^{-1}$ mode is observed to disappear while the 1590~cm$^{-1}$
mode remains.  Additional modes begin to become more visible such as
the 1265~cm$^{-1}$ and 1480~cm$^{-1}$ modes seen in Fig S3C.  At even
lower concentrations such as 1 nM, the $p$MBA signal is again more
intense with even more blinking as seen in Fig. S3F (which has been
plotted with intensity on a log scale).  The 1077~cm$^{-1}$ mode is
again unseen while the 1590~cm$^{-1}$ mode begins to fluctuate in
intensity even more.  The blinking becomes much more intense with the
intensity of the signal periodically reaching close to ten times the
maximum intensity observed for $p$MBA at 1~$\mu$M.  We suggest that
the increased blinking and larger amplitude signals are a result of
the molecules not being as tightly packed on the surface in the
1~$\mu$M and 1~nM cases as in the 1~mM case.  As a result of looser
packing, the molecules are free to explore more surface conformations,
including those with more and different charge transfer with the Au
surface.

We point out that these $p$MBA spectra are distinct from those seen in
physisorbed carbon contamination on initally clean junctions.  These
data persist at high incident powers and do not show ``arrival''
phenomena as described in the subsequent section.  Furthermore, they
are unlikely to originate from photodecomposition of $p$MBA, since the
illumination conditions are identical for all coverages.

\begin{figure}[h!]
\begin{center}
\includegraphics[clip, width=12cm]{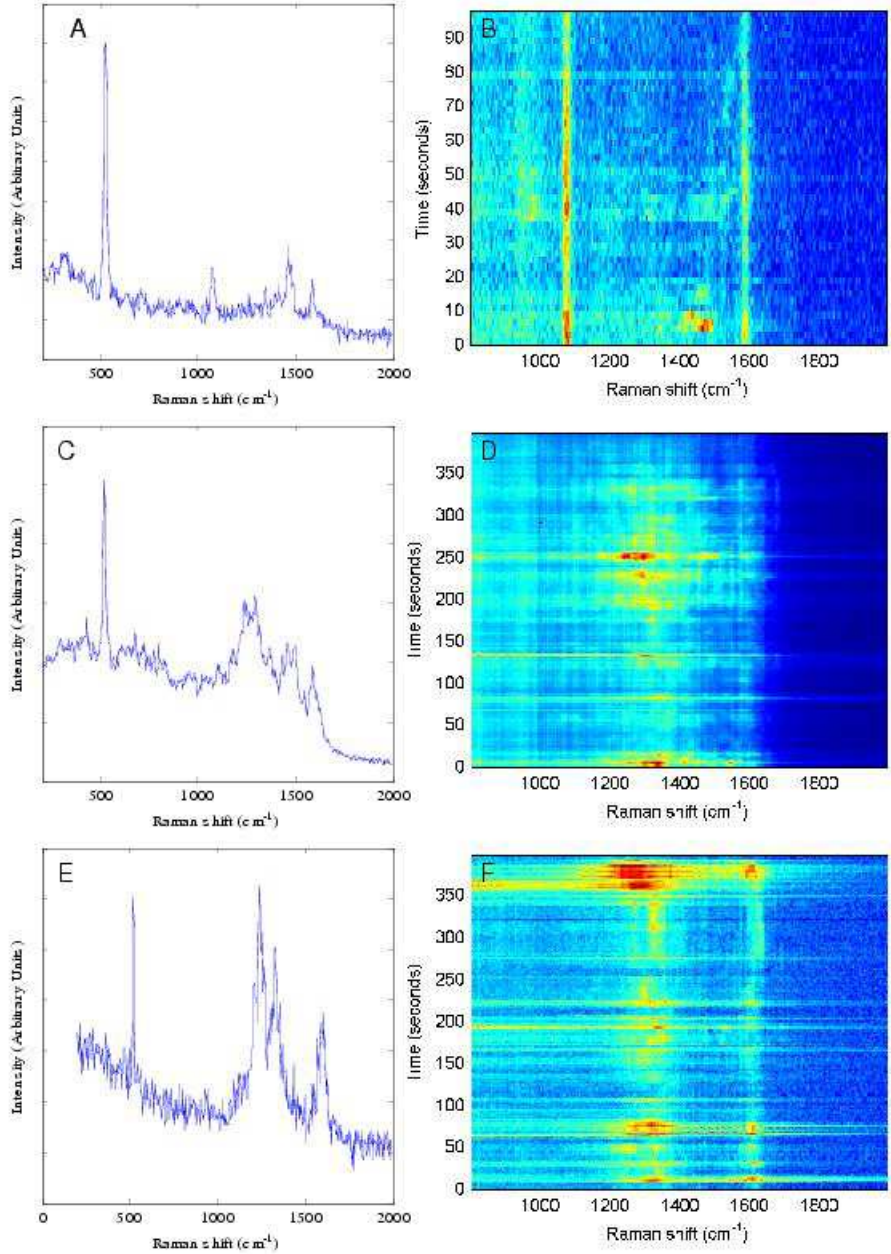}
\end{center}
%\vspace{-5mm}
\caption{\small 
(A) Raman from $p$MBA  at 1~mM concentration taken at $t=10$~s.
(B) Corresponding time spectrum for 1~mM.
(C) Raman from $p$MBA at 1~$\mu$M concentration taken at $t=251$~s.
(D) Corresponding time spectrum for 1~$\mu$M.
(E) Raman from $p$MBA at 1~nM concentration taken at $t = 24.5$~s
(F) Corresponding time spectrum for 1~nm, plotted on log intensity scale.
}
\label{suppfig3}
%\vspace{-5mm}
\end{figure}

\clearpage

\section{Detection of adsorbed contaminants}

Due to the large enhancements possible with the nanogaps, contamination from airborne absorbates occurs readily in the absence of assembled molecules on the nanogap surface.  We have observed the absorption of contaminants onto the surface of clean nanogaps in as little as 10 minutes.  Collecting Raman spectra every 4 seconds, we can observe the appearance of contaminants on the surface as seen in Fig. S4A and S4B.  It is difficult to identify the contaminants, as the spectra observed have large variations, although carbon ring modes are often observed in conjunction with other modes.  Furthermore the Raman signal from contaminants often blinks {\it very} strongly, with periods of no or weak signal followed by several seconds of intense blinking, as seen in Fig S4C.  The changes in intensity can be more than a factor of 100.  Again we suggest that the strong blinking is a result of the weak attachment of the contaminants to the nanogap surface, allowing them to move considerably and explore many interactions with the Au surface.  As previously mentioned, these contamination spectra are {\it not} observed when molecules of interest have been preassembled deliberately on the electrode surface.  The likely explanation for this is that the self-assembled later sterically prevents contaminants from arriving at the nanogap region of maximum field enhancement.

\begin{figure}[h!]
\begin{center}
\includegraphics[clip, width=12cm]{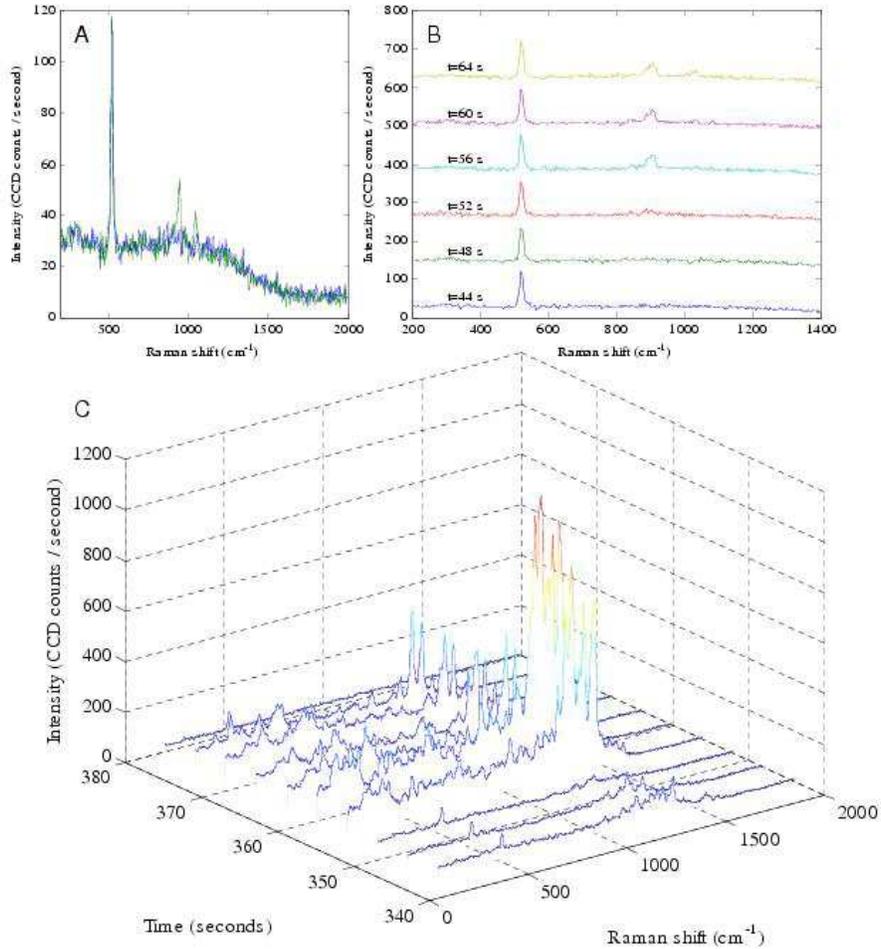}
\end{center}
%\vspace{-5mm}
\caption{\small 
(A) Raman spectra for clean bowtie (blue) and clean bowtie after a few minutes exposed to the air (green).  This change in the Raman spectrum is indicative of contamination for surface absorbed molecules from the air.
(B) Raman spectra for a clean bowtie showing the onset of a contaminant signal at 900~cm$^{-1}$ as time progresses.
(C) Waterfall plot showing the {\it extremely} strong blinking observed for adsorbed contamination.  The fluctuations are much larger than the those observed for dense coverage of $p$MA, $p$MBA, or P3HT.  Notice the scale relative to the 520~cm$^{-1}$ Si peak seen at $t=340$~s.
}
\label{suppfig4}
%\vspace{-5mm}
\end{figure}

\clearpage

\section{FDTD calculations}

The optical properties of the bowtie structure were calculated using
the Finite-Difference Time-Domain method (FDTD) using a Drude
dielectric function with parameters fitted to the experimental data
for gold. This fit provides an accurate description of the optical
properties of gold for wavelengths larger than
500~nm [S1].  These calculations do not account
for reduced carrier mean free path due to surface scattering in the
metal film, nor do they include interelectrode tunneling.  However,
such effects are unlikely to change the results significantly.

The bowtie is modeled as a two finite triangular structures as
illustrated in Fig.~4A of the manuscript. Our computational method requires the
nanostructures to be modeled to be of finite extent. The plasmon modes
of a finite system are standing modes with frequencies determined by
the size of the sample and the number of nodes of the surface charge
distribution associated with the plasmon. For an extended system such
as the bowties manufactured in this study, the plasmon resonances can
be characterized as traveling surface waves with a continuous
distribution of wavevectors.

A series of calculations of bowties with increasing length reveals
that the optical spectrum is characterized by increasingly densely
spaced plasmon resonances in the wavelength regime 500-1000~nm and a
low energy finite-size induced split-off state involving plasmons
localized on the outer surfaces of the bowtie. For a large bowtie, we
expect the plasmon resonances in the 500-1000~nm wavelength interval
to form a continuous band [S2].

The electric field enhancements across the bowtie junction for the
plasmon modes within this band are relatively similar with large and
uniform enhancements in the range of 50-150. The magnitudes of the
field enhancements were found to increase with increasing size of the
bowtie structure.  For instance, the maximum field enhancement factor was
found to be 115 for a 200~nm bowtie (Each half of the bowtie is 
modeled as a truncated triangle 200~nm long.) and 175 for a 400~nm bowtie. Our
use of a finite gridsize also underestimates the electric field
enhancements[S3]. Thus our calculated electric
field enhancements are likely to {\it significantly underestimate} the
actual electric field enhancements in the experimentally manufactured
bowties.

For a perfectly symmetric bowtie,
significant field enhancements are only induced for incident light
polarized across the junction.  If the mirror symmetry is broken, for
instance by making one of the structures thicker or triangular, large
field enhancements are induced for all polarizations of incident
light.

To investigate the effects of nanoasperities, FDTD calculations were
performed for a bowtie with two semi-spherical protrusions in the
junction as shown in Fig.~4 of the main text, and Figs. S5-S7 of the
Supporting Online Material. As expected, the presence of these
protrusions does not influence the optical spectrum. However, the
local field enhancements around the protrusions become very large,
typically three or four times higher than for the corresponding
structure without the defect. The physical mechanism for this increase
is an antenna effect caused by the coupling of plasmons localized on
the protrusion with the extended plasmons on the remaining bowtie
structure [S4].

\begin{figure}[h!]
\begin{center}
\includegraphics[clip, width=16cm]{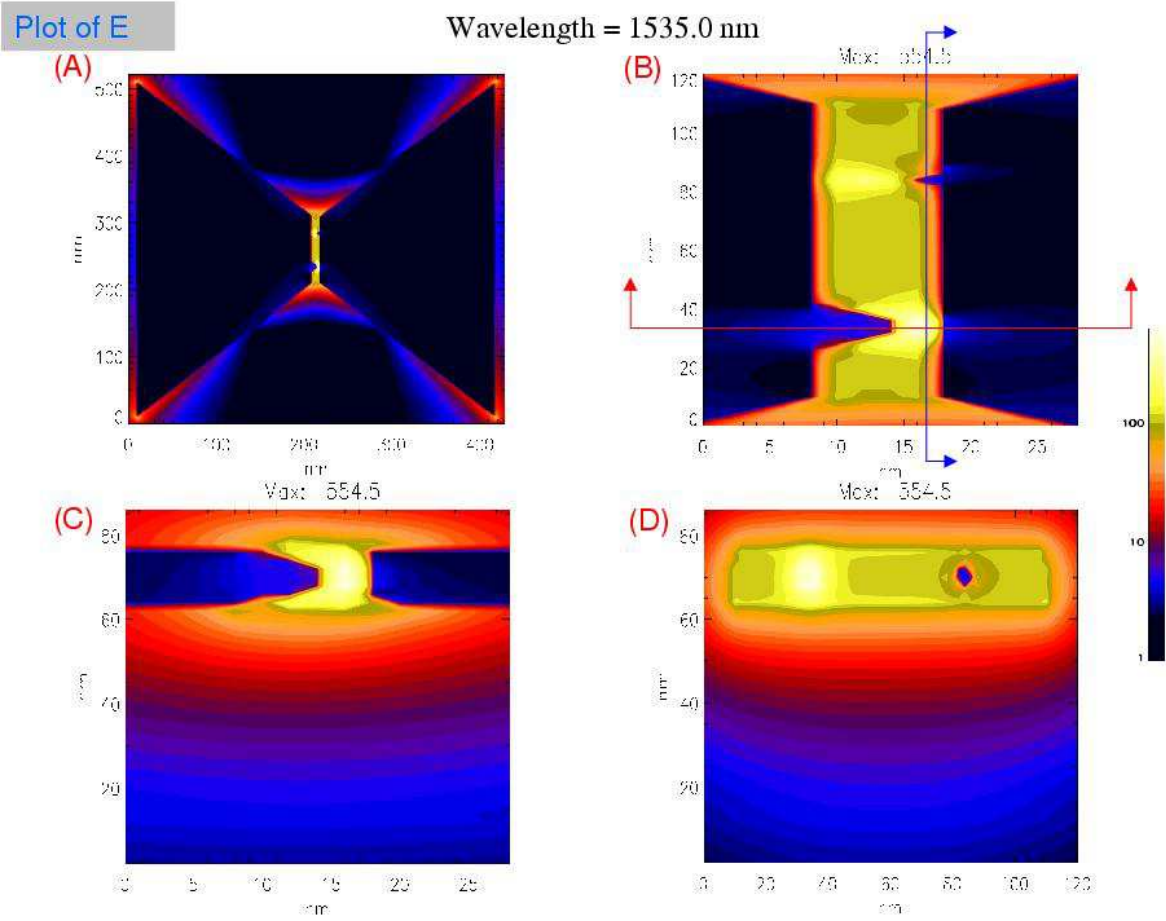}
\end{center}
%\vspace{-5mm}
\caption{\small Maps of FDTD-calculated $|{\mathbf E}|$ for the
1535~nm mode indicated in the main manuscript's Fig.~4A.  Color
scale is logarithmic in $|{\mathbf E}|/|{\mathbf E}_{\mathrm inc}|$.
Illumination direction is normal incidence, with electric field polarization
oriented horizontally in (A)-(C).  Maximum {\it field} enhancements
are shown.
(A) Overall view.  (B) Close-up of interelectrode gap showing 
asperities.  (C) Side-view of section indicated in (B) in red.
(D) Side view of section indicated in (B) in blue.
}
\label{suppfig5}
%\vspace{-5mm}
\end{figure}

\begin{figure}[h!]
\begin{center}
\includegraphics[clip, width=16cm]{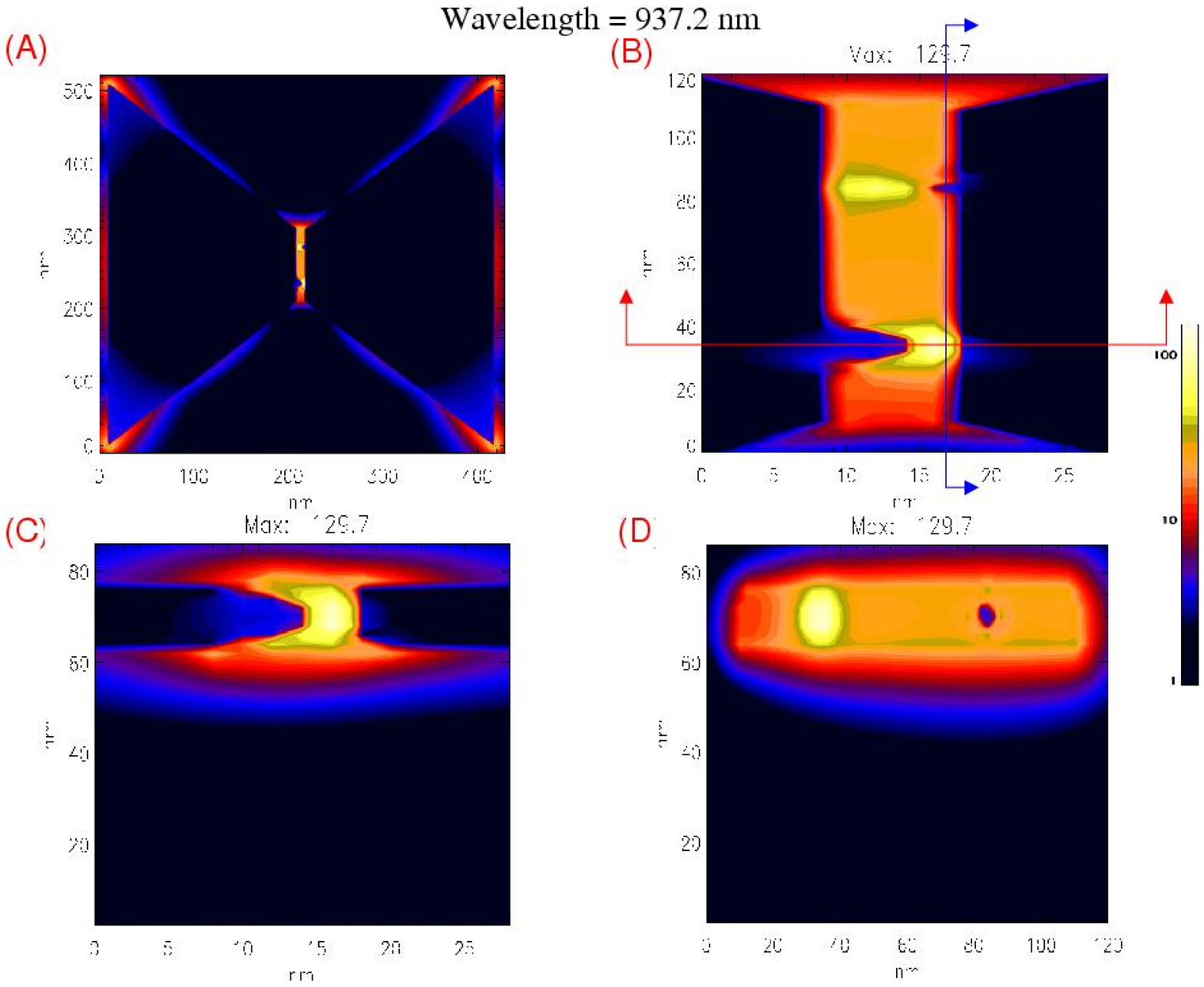}
\end{center}
%\vspace{-5mm}
\caption{\small Maps of FDTD-calculated $|{\mathbf E}|$ for the
937~nm mode indicated in the main manuscript's Fig.~4A.  Color
scale is logarithmic in $|{\mathbf E}|/|{\mathbf E}_{\mathrm inc}|$.
Illumination direction is normal incidence, with electric field polarization
oriented horizontally in (A)-(C).  Maximum field enhancements are shown.
(A) Overall view.  (B) Close-up of interelectrode gap showing 
asperities.  (C) Side-view of section indicated in (B) in red.
(D) Side view of section indicated in (B) in blue.
}
\label{suppfig6}
%\vspace{-5mm}
\end{figure}

\begin{figure}[h!]
\begin{center}
\includegraphics[clip, width=16cm]{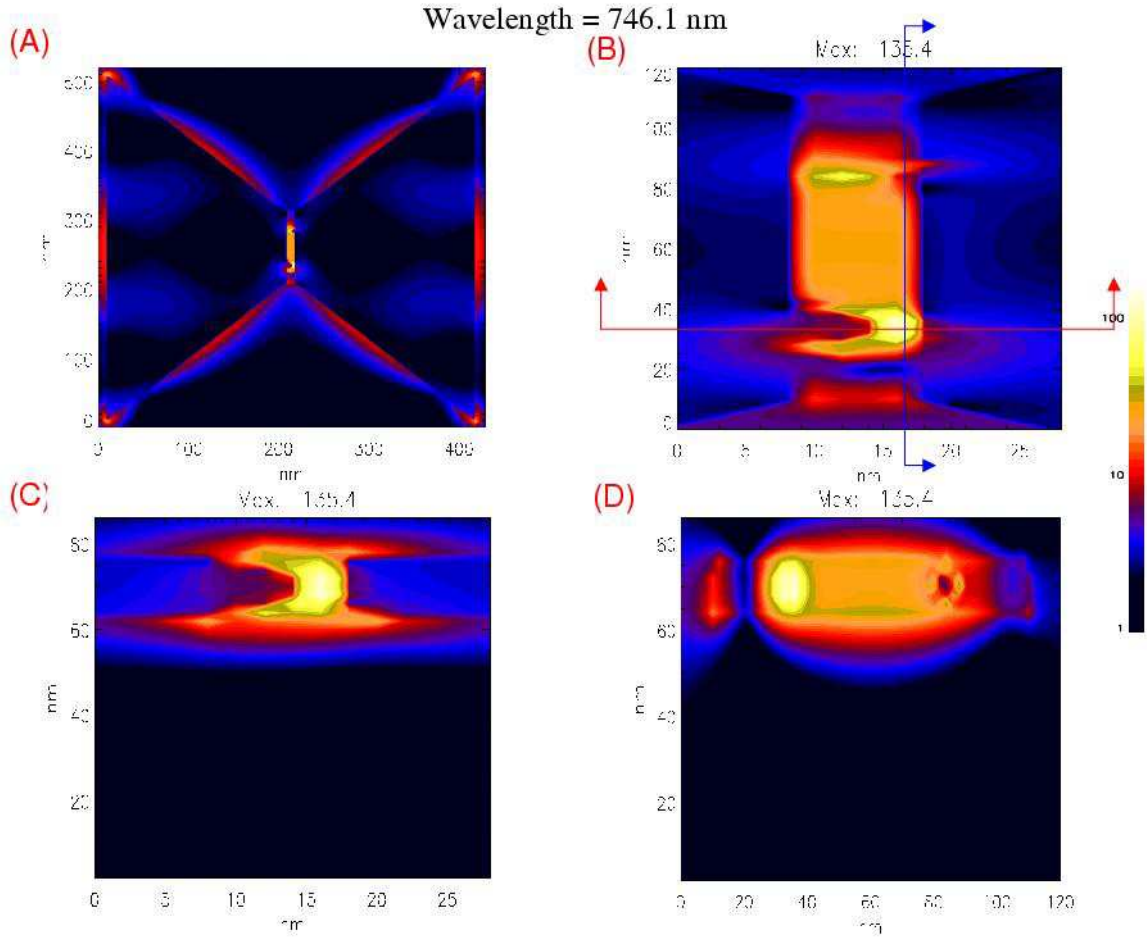}
\end{center}
%\vspace{-5mm}
\caption{\small Maps of FDTD-calculated $|{\mathbf E}|$ for the
746~nm mode indicated in the main manuscript's Fig.~4A.  Color
scale is logarithmic in $|{\mathbf E}|/|{\mathbf E}_{\mathrm inc}|$.
Illumination direction is normal incidence, with electric field polarization
oriented horizontally in (A)-(C).  Maximum field enhancements are shown.
(A) Overall view.  (B) Close-up of interelectrode gap showing 
asperities.  (C) Side-view of section indicated in (B) in red.
(D) Side view of section indicated in (B) in blue.
}
\label{suppfig7}
%\vspace{-5mm}
\end{figure}

\clearpage

\section{Enhancement estimate}

To estimate an enhancement based on the data of Fig.~3 in the main
text, it was necessary to understand the effective count rate per
molecule of Raman scattering from bulk $p$MA in our measurement setup.
This requires knowing the effective volume probed by the WITec
system when the laser is focused on a bulk $p$MA crystal.

The full-width-half-maximum (FWHM) of the laser spot size was found to
be 575~nm.  This was determined by measuring the count rate of the
Rayleigh scattering peak (at zero wavenumbers) as a function of
position as the beam was scanned over the edge of a Au film on a Si
substrate.  Averaging 16 such scans, the Rayleigh intensity was fit to
the form of an integrated gaussian to determine the FWHM of the
gaussian beam.  The 575~nm figure is likely an overestimate due to
systematic noise in the flat regions of the fit.

For a gaussian beam with intensity of the form $\propto
e^{-\frac{r^{2}}{2 \sigma^{2}}}$, the FWHM~$= 2 \sqrt{2 \ln 2}
\sigma$.  The effective radius of an equivalent cylindrical beam is $2
\sigma$, or 346~nm in this case.  The effective confocal
depth [S5] was determined by measuring the 520~cm$^{-1}$
Si Raman peak as a function of vertical displacement of a blank
substrate.  The effective depth profile was determined by numerical
integration of the Si data using matlab.  The effective volume probed
by the beam is $1.92 \times 10^{-12}$~cm$^{3}$.  From the bulk
properties of $p$MA, this corresponds to $1.09\times 10^{10}$
molecules.

The count rate for the bulk $p$MA 1077~cm$^{-1}$ line, corrected by
the ratio of (Si SERS rate/Si bulk rate) to accomodate for the
difference in laser powers, is 46 counts/s, compared with 203 counts/s
for the SERS data of Fig.~3.  This leads to the enhancement estimate
quoted in the main text of $5 \times 10^{8}$.

%\begin{thebibliography}{10}

%\bibitem{SOubreNordlander04JPCB}
\noindent [S1] Oubre, C.; Nordlander, P. {\it J. Phys. Chem. B\/} {\bf 108}, {\it 108},  17740-17747.

%\bibitem{SNordlanderLe06APB}
\noindent [S2] Nordlander, P.; Le, F. {\it Appl. Phys. B\/} {\bf 2006}, {\it 84}, 35-41.

%\bibitem{SOubreNordlander05JPCB}
\noindent [S3] Oubre, C.; Nordlander, P. {\it J. Phys. Chem. B\/} {\bf 2005}, {\it 109}, 10042-10051.

%\bibitem{SHaoetAl07NL}
\noindent[S4] Hao, F.; Nehl, C.~L.; Hafner, J.~H.; Nordlander, P. {\it Nano Lett.\/} {\bf 2007}, {\it 7}, 10.1021/nl062969c.

%\bibitem{SCaietAl98SS}
\noindent [S5] Cai, W.B.; Ren, B.; Li, X.~Q.; Shi, C.~X.; Liu, F.~M.; Cai, X.~W.; Tian, Z.~Q. {\it Surf. Sci.\/} {\bf 1998}, {\it 406}, 9-22.

%\end{thebibliography}

\end{document}